\documentclass{jfm}
\usepackage{commath}
\usepackage{graphicx}
\usepackage{newtxtext}
\usepackage{newtxmath}
\usepackage{natbib}
\usepackage{hyperref}
\usepackage{comment}
\usepackage{color,soul}
\hypersetup{
    colorlinks = true,
    urlcolor   = blue,
    citecolor  = black,
}

\newcommand{\RomanNumeralCaps}[1]
\linenumbers

\title{Fate of bubble clusters rising in a quiescent liquid} 

\author{
Tian Ma\aff{1,}$^{*,}$\corresp{\email{tian.ma@hzdr.de}},
Hendrik Hessenkemper\aff{1,}$^{*,}$\corresp{\email{h.hessenkemper@hzdr.de}},
Dirk Lucas\aff{1}
\and Andrew D. Bragg\aff{2}\corresp{\email{andrew.bragg@duke.edu}}
}

\affiliation{\aff{1}Helmholtz-Zentrum Dresden -- Rossendorf, Institute of Fluid Dynamics, 01328 Dresden, Germany
\aff{2}Department of Civil and Environmental Engineering, Duke University, Durham, NC 27708, USA}

\begin{document}
\maketitle
\def\thefootnote{*}\footnotetext{These authors contributed equally to this work}

\begin{abstract}
We use experiments to study the evolution of bubble clusters in a swarm of freely rising, deformable bubbles. A new machine learning-aided algorithm allows us to identify and track bubbles in clusters and measure the cluster lifetimes. The results indicate that contamination in the carrier liquid can enhance the formation of bubble clusters and prolong the cluster lifetimes. The mean bubble rise velocities conditioned on the bubble cluster size are also explored, and we find a positive correlation between the cluster size and the rise speed of the bubbles in the cluster, with clustered bubbles rising up to $20\%$ faster than unclustered bubbles.
\end{abstract}

\section{Introduction}\label{sec: introduction}

Bubbles rising though a liquid is a frequently occurring situation in nature (e.g. bubble plumes rising from the bottom of a lake), daily life (e.g. bubble chains rising in carbonated drinks) and technology (e.g. waste water treatment). For the case of bubble swarms rising freely in a quiescent fluid, the bubbles tend to distribute inhomogeneously, spontaneously forming clusters.  When the bubbles are large enough, the rising bubbles generate turbulence in the liquid, a phenomena referred to as bubble-induced turbulence (BIT). This BIT in turn influences the clusters and understanding this is important both for its own sake, and also due to its impact on many aspects of bubble motion, including their rise velocities, collisions, dispersion and the intensity of the BIT generated \citep{2008_Takagi,2013_Tagawa,2018_Lohse,2020_Mathai}. %Note that in the context of swarms, bubble clustering can be quantified by comparing their distributional properties relative to that which they would have if they were uniformly distributed, similar to the case for inertial particles in turbulent flows \citep{22_brandt}.

Early numerical investigations \citep[e.g.][]{1993_Smereka} on bubble clustering in the presence of BIT assumed spherical bubbles and considered potential flow, and found that the bubbles form large horizontal clusters as they rise. However, subsequent experiments and more realistic numerical simulations did not clearly observe such clusters, but 2-bubble clusters with a wide range of orientations were identified as the most commonly occurring clusters \citep{2001_Zenit,2003_Bunner,2005_Esmaeeli}. The prevalence of 2-bubble clusters has motivated the community to explore the dynamics of bubble pairs with varying separations and orientations, usually for the case of clean bubbles \citep{2011_Hallez}. The two extreme cases are bubbles aligned side-by-side \citep{2003_Legendre} and in-line \citep{2021_Zhang}. The results indicate that the side-by-side configuration is more stable than the in-line case, and this is because for the in-line configuration a slight transverse movement of the trailing bubble relative to the leading bubble makes it \textquoteleft feel\textquoteright\ a shear flow, that can drive the trailing bubble out of the leading bubbles wake. Nevertheless, stable in-line bubble chains are often observed in carbonated drinks. This contradiction was recently explained by \cite{2022_Atasi} as being due to the combined effects of bubble deformation and contamination in such liquids that can result in a reversal of the lift force and stable chain. In addition to exploring the stability of nearby bubble pairs, the impact of neighbouring bubbles on their rise velocity has been investigated and compared to the case of isolated bubbles. \cite{2011_Hallez} showed that the side-by-side configuration maximizes the drag force acting on a pair of bubbles, while the in-line bubble configuration minimizes the drag due to wake entrainment for the trailing bubble. 

These studies on bubble pair dynamics have provided much insight, however, there are many open questions concerning the behaviour of bubble swarms where two or more bubbles may be clustered together, whose motion may also be affected by the wakes of other bubbles and bubble clusters in the flow. Indeed, while the rise velocity of bubble pairs in a quiescent liquid is well understood, its behaviour in the context of bubble swarms is debated. For example, the experiments of \cite{1995_Stewart} and \cite{1999_Bruecker} for large deformable bubbles in a swarm found that the mean rise velocity was considerably larger than that for a single bubble. However, this contradicts other experimental \citep{1979_Ishii} and numerical \citep{2011a_Roghair} studies for large bubbles which argue that the mean bubble rise velocity decreases monotonically as the gas void fraction is increased.

Several fundamental questions remain mostly unexplored: what is the probability to form clusters involving $N_b$ number of bubbles? What is the lifetime of these clusters? How does the rise velocity of bubbles in a cluster depend on $N_b$? How do the answers to these questions depend on contaminants in the liquid? In this paper we explore these questions experimentally by tracking thousands of deformable bubbles in a vertical column, using a recently developed machine-learning algorithm to detect and follow the evolution of bubble clusters, and explore how the bubble rise velocities depend on $N_b$. We also consider the effect of surfactants to provide a more complete picture for real systems where contaminants may cause behavior that differs substantially from that of an idealized clean system.

\section{Experimental method}\label{sec: Experiment}

\subsection{Experimental set-up}

The experimental apparatus is identical to that in \cite{2022_Ma}, and we therefore refer the reader to that paper for additional details; here, we summarize. The experiments were conducted in a rectangular bubble column (depth $50\,\mathrm{mm}$ and width $112.5\,\mathrm{mm}$), with a water fill height of 1,000 mm. Air bubbles are injected through 11 spargers which are homogeneously distributed at the bottom of the column. 

We use tap water in the present work as the base liquid and consider two different bubble sizes by using spargers with different inner diameters. For each bubble size, we manipulate the gas flow rate and ensure that all cases are not in the heterogeneous regime of dispersed bubbly flows. In total, we have six mono-dispersed cases (see supplementary movies 1-6) labelled as \textit{SmTapLess}, \textit{SmTap}, \textit{SmPen+}, \textit{LaTapLess}, \textit{LaTap}, and \textit{LaPen+} in table \ref{tab: bubble para}, including some basic characteristic dimensionless numbers for the bubbles. Here, \textquotedblleft\textit{Sm}/\textit{La}\textquotedblright\ stand for smaller/larger bubbles, \textquotedblleft\textit{Pen+}\textquotedblright\ stands for corresponding cases added by 1,000 ppm 1-Pentanol, and \textquotedblleft\textit{Less}\textquotedblright\ stands for lower gas void fraction than \textasteriskcentered\textasteriskcentered\textit{Tap}/\textasteriskcentered\textasteriskcentered\textit{Pen+} cases for smaller/larger bubbles, respectively. It should be noted that the three cases with larger bubble sizes have higher gas void fractions than the three cases with smaller bubbles. This is because in our setup it is not possible to have the same flow rate for two different spargers while also maintaining a homogeneous gas distribution for mono-dispersed bubbles. Furthermore, the bubble size is slightly reduced when adding 1-Pentanol for both types of sparger. This is due to the influence of the surfactants that reduce the surface tension and hence affect the bubble formation at the rigid orifice.

To identify and track bubble clusters, we use planar shadow images obtained by recording the flow with a high-speed camera and illuminating the setup with a LED. The measurement resolution in time and space are 250 fps and 59.9 µm/Px, respectively, with a field of view (FOV) of $90\,\mathrm{mm}\times76\,\mathrm{mm}$. For each case, we record 1,000 sequences with each having 70-75 frames -- approximately the time a bubble passing through the complete image height. 

\begin{table}
	\begin{center}
		\def~{\hphantom{0}}
		\begin{tabular}{ccccccc}
			Parameter  
			&\textsl{SmTapLess}&\textsl{SmTap}&\textsl{SmPen+}&\textsl{LaTapLess}&\textsl{LaTap}&\textsl{LaPen+}\\	
			\hline
			$\alpha$   &0.51\%&0.79\%&0.71\%&1.2\%&1.98\%&1.91\%\\
			$d_b \;(\mathrm{mm})$    &3&3.1&2.7&4&4.3&3.8\\
			$\chi$     &1.9&1.9&1.2&1.9&2.0&1.3\\
			$L_b \;(\mathrm{mm})$    &14.1&12.7&11.4&14.1&13.0&11.2\\
			%$\dot{V} \;(\mathrm{L/min})$  &0.44&0.62&0.44&1&1.4&1\\
			$Ga$       &512&538&437&788&879&730\\
			$Eo$       &1.29&1.38&1.05&2.30&2.66&2.08\\
			$Re_b$     &755&739&493&912&1022&782\\
			$C_D$      &0.61&0.70&1.04&0.98&0.97&1.14\\
		\end{tabular}
		\caption{Selected characteristics of the six bubble swarm cases. Here, $\alpha$ is the averaged gas void fraction, $d_b$ the equivalent bubble diameter, $\chi$ the aspect ratio, $L_b$ the inter-bubble distance, $\dot{V}$ the gas flow rate at injection, $Ga\equiv\sqrt{\left|\pi_\rho-1\right|gd_{b}^{3}}/\nu$ the Galileo number, $Eo\equiv\Delta\rho gd_b^2/\sigma$, the E\"{o}tv\"{o}s number. The bubble Reynolds number $Re_b$ and  drag coefficient $C_D$ are based on $d_b$ and the bubble to fluid relative velocity.} \label{tab: bubble para}
	\end{center}
\end{table}

\subsection{Bubble identification and tracking}

In our study only one camera is used, however, as will be shown in \S\,\ref{subsec}, we are nevertheless able to perform quasi-3D tracking of bubbles. Independent on the number of cameras used, the task to track bubbles in image sequences can be done in a detect-to-track or in a track-to-detect fashion. While the former links previously detected bubbles in each frame to form suitable tracks, the latter uses extrapolations of already established tracks to detect bubbles in follow-up images. We use the former detect-to-track strategy, which allows to incorporate detections among multiple frames to establish tracks, with however relying more strongly on an accurate detector that finds bubbles in individual frames.

\begin{figure}
	\centering
	\includegraphics[width=12.7cm]{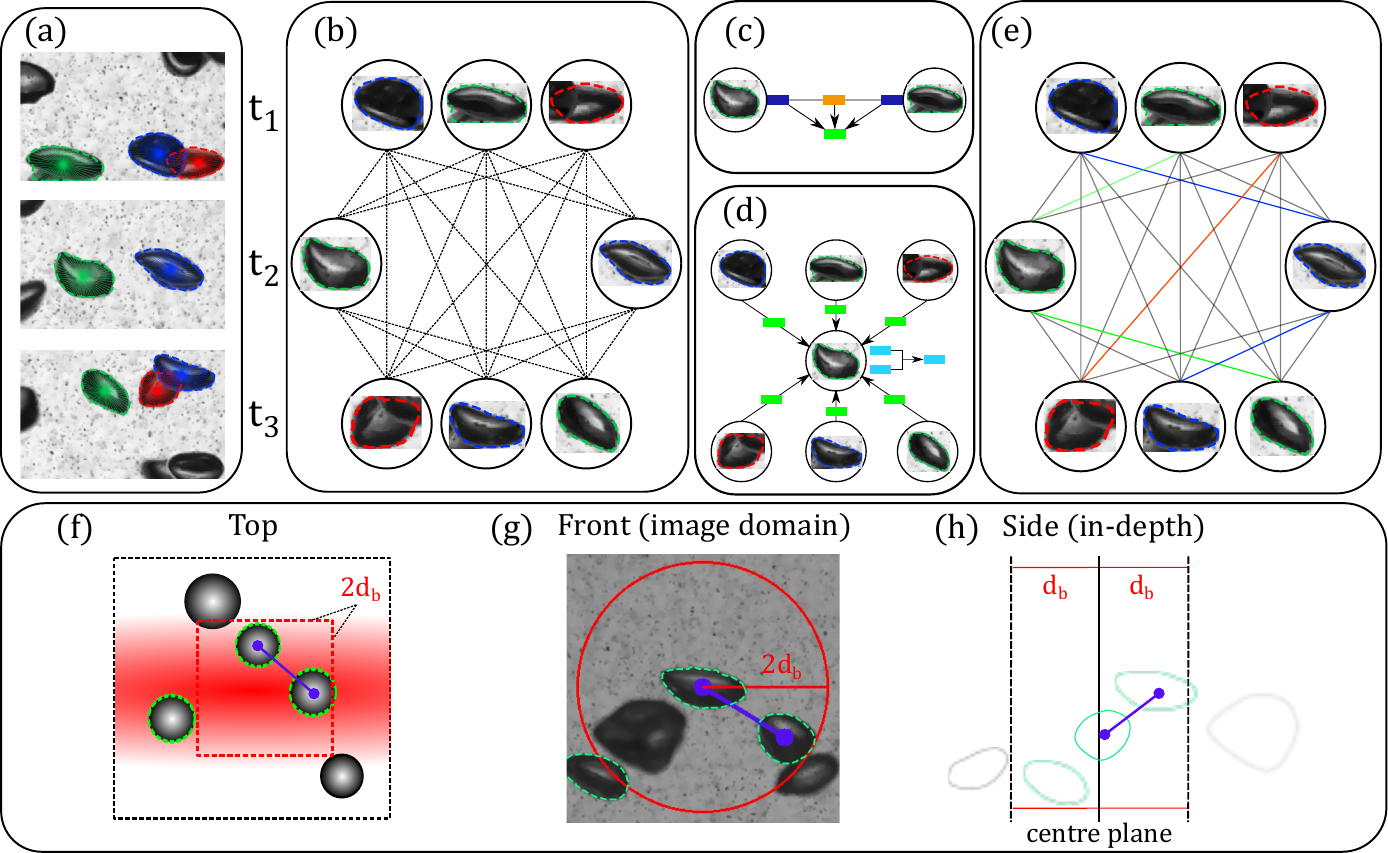}
	\caption{Steps of the tracking algorithm: (\textit{a}) detections represented with 64 radial vectors, (\textit{b}) graph construction, (\textit{c}) feature encoding with node encoder MLP (dark blue rectangle) and edge encoder MLP (orange rectangle) together with edge update MLP (green rectangle), (\textit{d}) time-aware node update MLPs (light blue rectangle), (\textit{e}) predicted active edges. Cluster search region: view from (\textit{f}) the top, (\textit{g}) the front and (\textit{h}) the side. Note that (\textit{f,h}) are only schematic representations, showing possible bubble arrangements.}
	\label{fig: ML}
\end{figure}

Even for low gas volume fractions, detecting bubbles in individual images is a challenging task since bubbles can overlap in the images. Fully overlapping bubbles cannot be detected, but partially overlapping bubbles can be dealt with and deep-learning-based strategies for this have recently shown very promising results \citep[e.g.][]{2021_Kim}. In our previous work \citep{2022_Hessenkemper}, we developed such an approach that used a trained convolutional neural network (CNN) to segment overlapping bubbles. Furthermore, the contour of each detected bubble is reconstructed using 64 radial vectors pointing from the segmentation centre to the boundary (figure \ref{fig: ML}\textit{a}), and the radial vectors of partly occluded bubbles are corrected using an additional multi-layer perceptron (MLP).

The subsequent tracking of multiple detected bubbles in close proximity poses further challenges, as the tracker not only has to be robust against inaccuracies of the detector, i.e. missing or false detections, but also has to be able to track bubbles that are fully occluded even for multiple times steps, while at the same time having numerous possible associations in the near vicinity. To solve these issues, a graph-based tracking formalism is used. Specifically, we follow the framework of \cite{2020_Braso}, utilizing multiple MLPs to predict valid connections of detections on graph structured data. The four main aspects of this tracking framework are described as follows. Details on network architectures, the created training dataset as well as validation tests are provided in the Supplementary Materials.

\textbf{Graph construction:} To track the bubbles, each sequence is modelled as a graph, with detections (bubbles) being the nodes of the graph and possible connections in time being the edges of the graph, i.e. a pair of detections forward or backward in time that are possibly from the same bubble  (figure \ref{fig: ML}\textit{b}). The task is then to classify the edges in active and non-actives edges, which at the end form a set of valid tracks that fulfil the so-called \textquoteleft flow conservation constraints\textquoteright -- each node having an active edge to at most one node forward in time and at most one node backward in time.

\textbf{Feature encoding:} For both the nodes and the edges of the graph, features are encoded with two separate MLPs  (figure \ref{fig: ML}\textit{c}). The node embeddings represent the appearance features of the detections, which are usually encoded with a CNN (\citealt{2020_Braso}). However, monochromatic bubble images show few distinct features, with the size and the shape of the bubble image being the most relevant ones. Thus, we have chosen the 64 radial vectors from the bubble detector as input for the node feature encoder, providing not only a more accurate description of the features bubble size/shape, but also a better 2D bubble contour in the case of overlapping bubbles due to the additional correction of the radial locations. The edge embeddings represent tracking-related features. For each detected pair of frames, the time increment, relative coordinate and size are fed into the edge feature encoding MLP to generate edge embeddings.

\textbf{Message Passing Network:} The core of the tracking algorithm is the Message Passing Network (MPN) whose main purpose is to update node and edge embeddings w.r.t. their surrounding nodes and edges in the graph, and this is done iteratively using message passing steps. First, the edge embeddings are updated by combining their embeddings with the embeddings of the adjacent pair of nodes and feeding them into an edge-update MLP  (figure \ref{fig: ML}\textit{c}). Then, a time-aware node update step is applied by aggregating edge embeddings of adjacent edges, which already contain information of connected nodes due to the previous edge update step. The time-awareness is introduced by at first aggregating and updating separately incoming edges, i.e. connections backward in time, and outgoing edges, i.e. connections forward in time, with individual MLPs and then concatenating the outcome to finally update the node embeddings with a node update MLP (figure \ref{fig: ML}\textit{d}). For each iteration, information of nodes one step further in time is passed through the network to the node/edge to be updated. Thus, the number of iterations defines the time increment of the information of other nodes that are supplied to the current node.

\textbf{Edge classification and post-processing:} After updating all node and edge embeddings with the MPN, the edges are classified with a classifier MLP (figure \ref{fig: ML}\textit{e}). The predictions are post-processed and remaining violations of the flow conservation constraints are treated with an exact rounding scheme (\citealt{2020_Braso}). Lastly, missing links in the trajectories are interpolated using bilinear interpolation and each trajectory is smoothed with a uniform filter.

\subsection{Identification and tracking of bubble clusters}\label{subsec}

The detection of bubble clusters at each time step follows a distance criteria between neighbouring bubbles whose centres in the 2D image domain are below a predefined threshold $2d_b$ from each individual case (figure \ref{fig: ML}\textit{g}). This value is mainly based on the work of \cite{2003_Legendre} that a considerable drag enhancement is observed for a bubble pair rising side-by-side within this distance. Tests for different thresholds ($2d_b\pm0.5d_b$) were conducted and the trends of the results in \S\,\ref{sec: Results} were found to be insensitive to the choice of this parameter.
Furthermore, since we attempt to detect the bubble cluster in a quasi-3D manner, we keep the in-focus region in the depth direction to also be $2d_b$ (figure \ref{fig: ML}\textit{f,h}). To estimate this depth distance to the centre plane we use the gray value gradient of the detected bubbles and consider only sharp bubbles in the shallow Depth of Field (DoF) region (see supplementary materials for more detail). In summary, we utilize a cylindric search volume, $\pi(2d_b)^2\times2d_b$, for the cluster identification, radially in the 2D image domain and linear in the depth direction. For all the cases, the mean inter-bubble distance $L_b$ based on the global void fraction (table \ref{tab: bubble para}) is much larger than the search radius $2d_b$, indicating that the bubble clusters to be discussed are dynamically significant.
\begin{figure}
	\centering
	\makebox[1.8em][l]{\raisebox{-\height}{(\textit{a})}}%
	\raisebox{-\height}{\includegraphics[height=2.4cm]{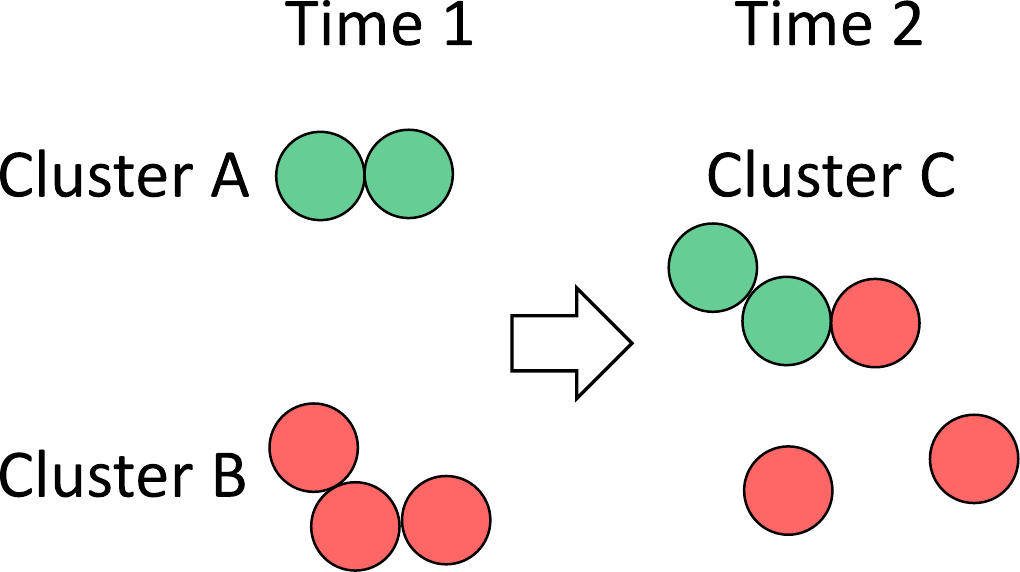}}
	\hspace{9mm}
	\makebox[2.0em][l]{\raisebox{-\height}{(\textit{b})}}%
	\raisebox{-\height}{\includegraphics[height=2.4cm]{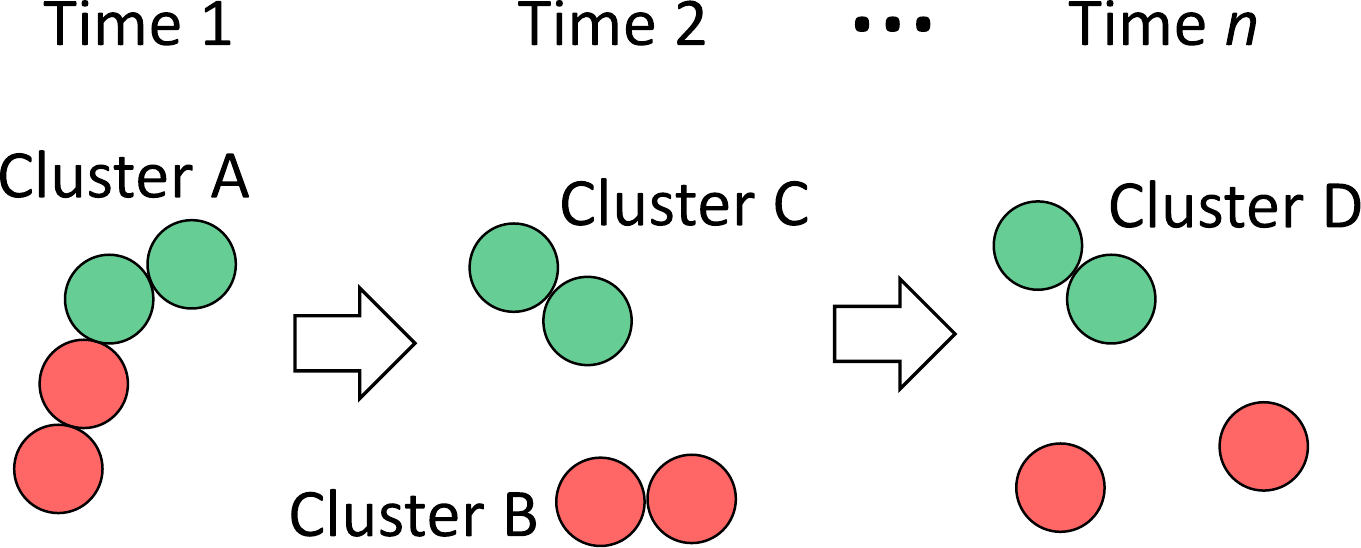}}
	\caption{Method of tracking clusters: (\textit{a}) illustrates an example of two clusters merging into one cluster; (\textit{b}) illustrates an example of one cluster separating into two clusters.} \label{fig: tracking}
\end{figure}

The cluster tracking strategy is inspired by the work of \cite{2020_Liu} for characterising the temporal evolution of inertial particle clusters in turbulence. Considering two clusters identified in two consecutive time steps ($\Delta t=1/250s$), we take both to be successive realizations of the same cluster if the number of bubbles they share is above a given threshold. The shared bubbles across clusters in successive time steps are termed connections. We consider forward-in-time and
backward-in-time connections, and apply thresholds on the fraction of connected bubbles over the total number of bubbles in each cluster. We illustrate in figure \ref{fig: tracking}(\textit{a}) an example: Cluster A (identified in time step 1) shares all its bubbles with cluster C (identified in time step 2), while C shares 2/3 of its bubbles with A. Therefore, the fractions of forward and backward connections between A and C are 1 and 2/3, respectively. On the other hand, B shares 1/3 of its bubbles with C, and C shares 1/3 of its bubbles with B. Thus, the forward and backward connections between B and C are 1/3 and 1/3, respectively.  Following \cite{2020_Liu}, two clusters in consecutive
time steps are identified as the same cluster when the fractions of their backward and forward connections are both $\geq1/2$. In the example of figure \ref{fig: tracking}(\textit{a}), A and C are recognized as belonging to the same cluster. The cluster lifetime is defined as the time elapsed between birth (the first instance a cluster is identified) and death (the last time it is recognized). Here, we explicitly include the lower threshold of $1/2$, since many 2-bubble clusters appear and require an additional criterion for tracking. In figure \ref{fig: tracking}(\textit{b}) we give an example where cluster A at time step 2 splits into B and C. To decide whether B or C should be regarded as the continuation of cluster A for the purposes of tracking, we consider whether cluster B or C persists longer into the future. In this example, while cluster C survives until time \textit{n}, B does not. Therefore, we regard C, D, and A as belonging to the same cluster, while cluster B is considered to be a newborn cluster at time step 2. This approach eliminates ambiguities since it ensures that a cluster at any instant can only be associated with at most one cluster either in the past or future.

\section{Results}\label{sec: Results}

\subsection{Probability to be clustered}\label{sec: probability}

\begin{figure}
\centering
\includegraphics[height=5.3cm]{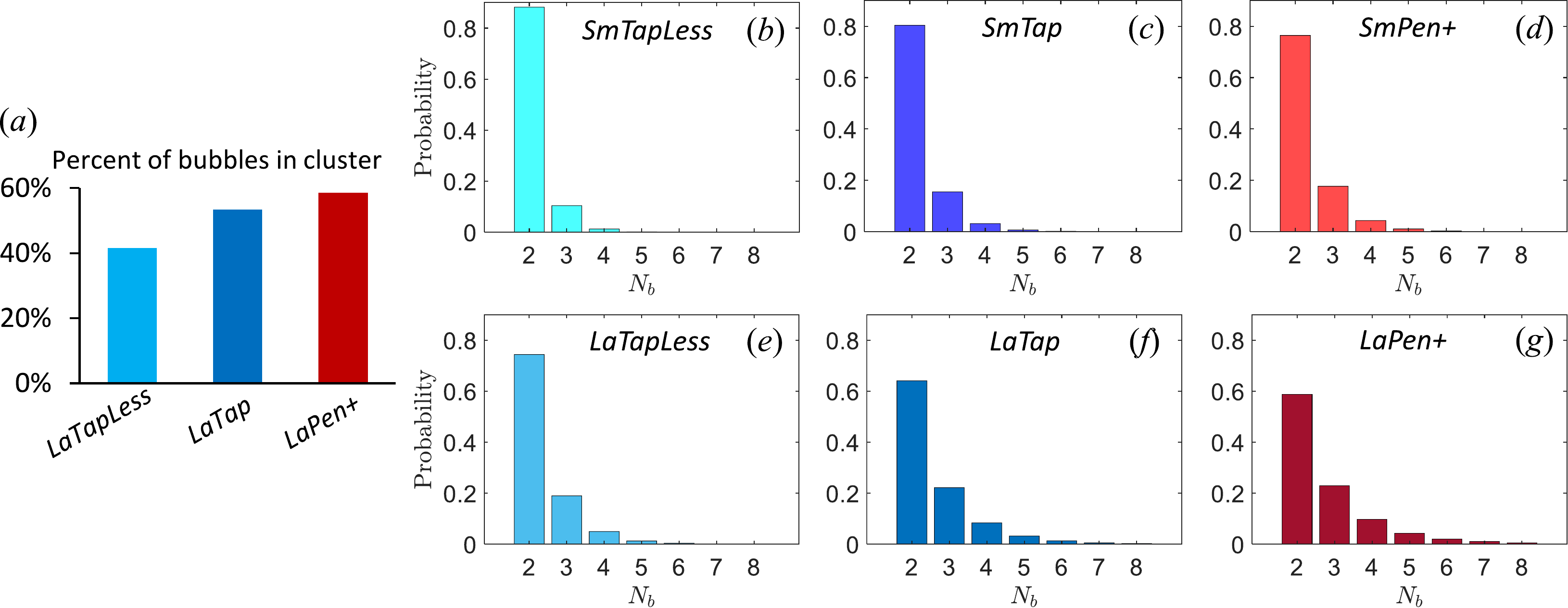}
\caption{(\textit{a}) Percentage of bubbles in cluster for 3 cases with larger bubbles. (\textit{b-g}) Probability of number of bubbles within a cluster for the different cases.} \label{fig: bub-in-clus}
\end{figure}

We first consider the percentage of bubbles in the flow that are clustered, and the results in figure \ref{fig: bub-in-clus}(\textit{a}) show that this percentage increases in the order of \textit{LaTapLess}, \textit{LaTap} to \textit{LaPen}+ for the larger bubbles. While the increase from \textit{LaTapLess} to \textit{LaTap} is quite understandable due to the increase of the gas void fraction, the result from \textit{LaPen}+ (whose $\alpha$ is sightly less than that of \textit{LaTap}) shows the surfactant promotes the formation of clusters. We obtained similar results (not shown) for the three cases with smaller bubbles. 

In figure \ref{fig: bub-in-clus}\textit{b-g} we consider the probability to find a given number of bubbles within a cluster for all cases, respectively. The results show that the probability decreases with increasing $N_b$, and consistent with previous studies, $N_b=2$ is the most common cluster size for all 6 cases \citep{2001_Zenit,2003_Bunner}. However, the results also show that $N_b=3,4$ clusters occur with non-negligible probability, and there are even rare events with $N_b=8$ clusters. The results also show that adding contaminants decreases the probability to form $N_b=2$ clusters, and increases the probability to form larger clusters, although the dependence is not too strong. For a fixed contaminant level, increasing $\alpha$ has the same effect.

\subsection{Lifetime}

\begin{figure}
	\begin{minipage}[b]{1.0\linewidth}
		\centering
		\makebox[1.2em][l]{\raisebox{-\height}{(\textit{a})}}%
		\raisebox{-\height}{\includegraphics[height=3.7cm]{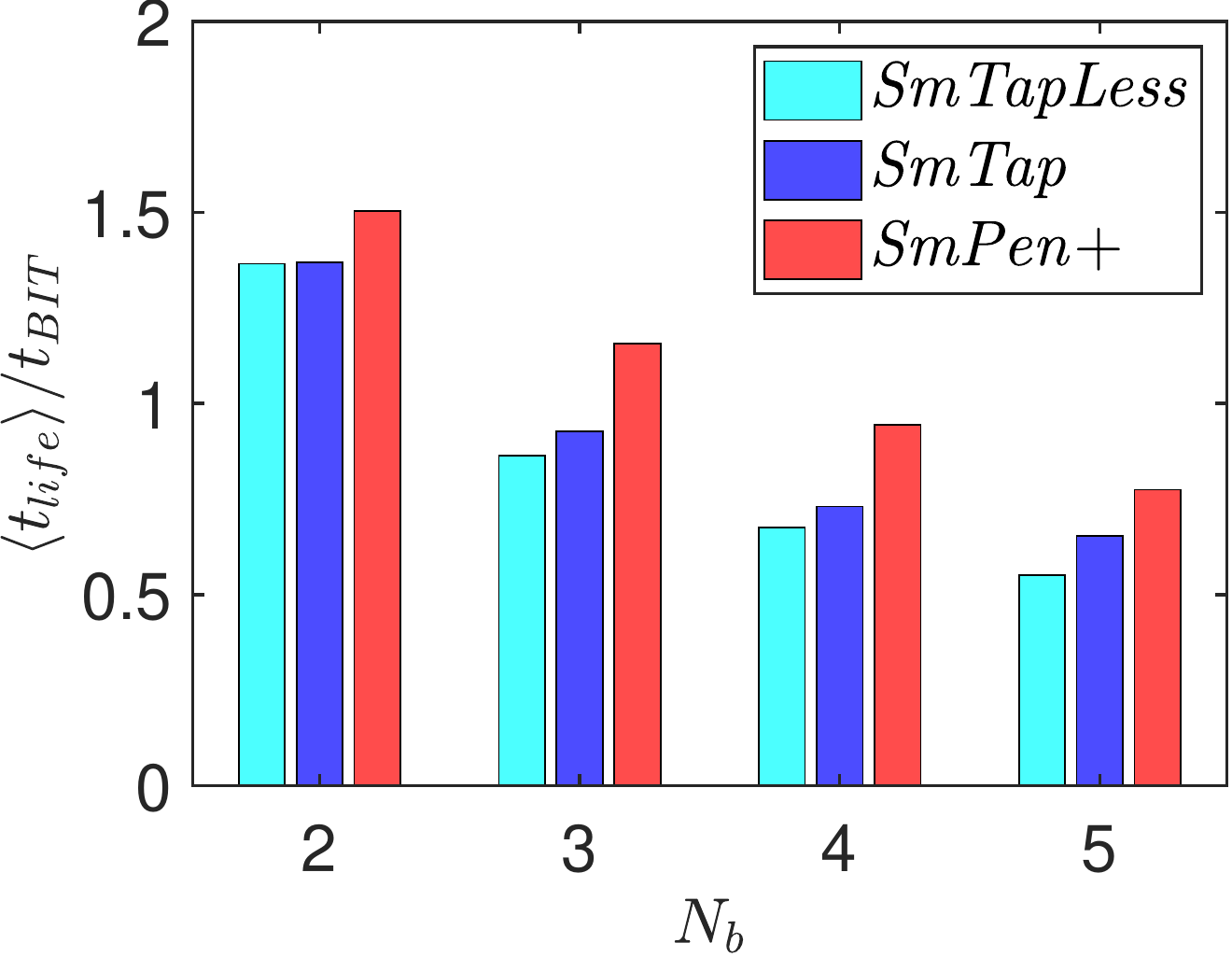}}
        \hspace{5mm}
		\makebox[1.2em][l]{\raisebox{-\height}{(\textit{b})}}%
		\raisebox{-\height}{\includegraphics[height=3.7cm]{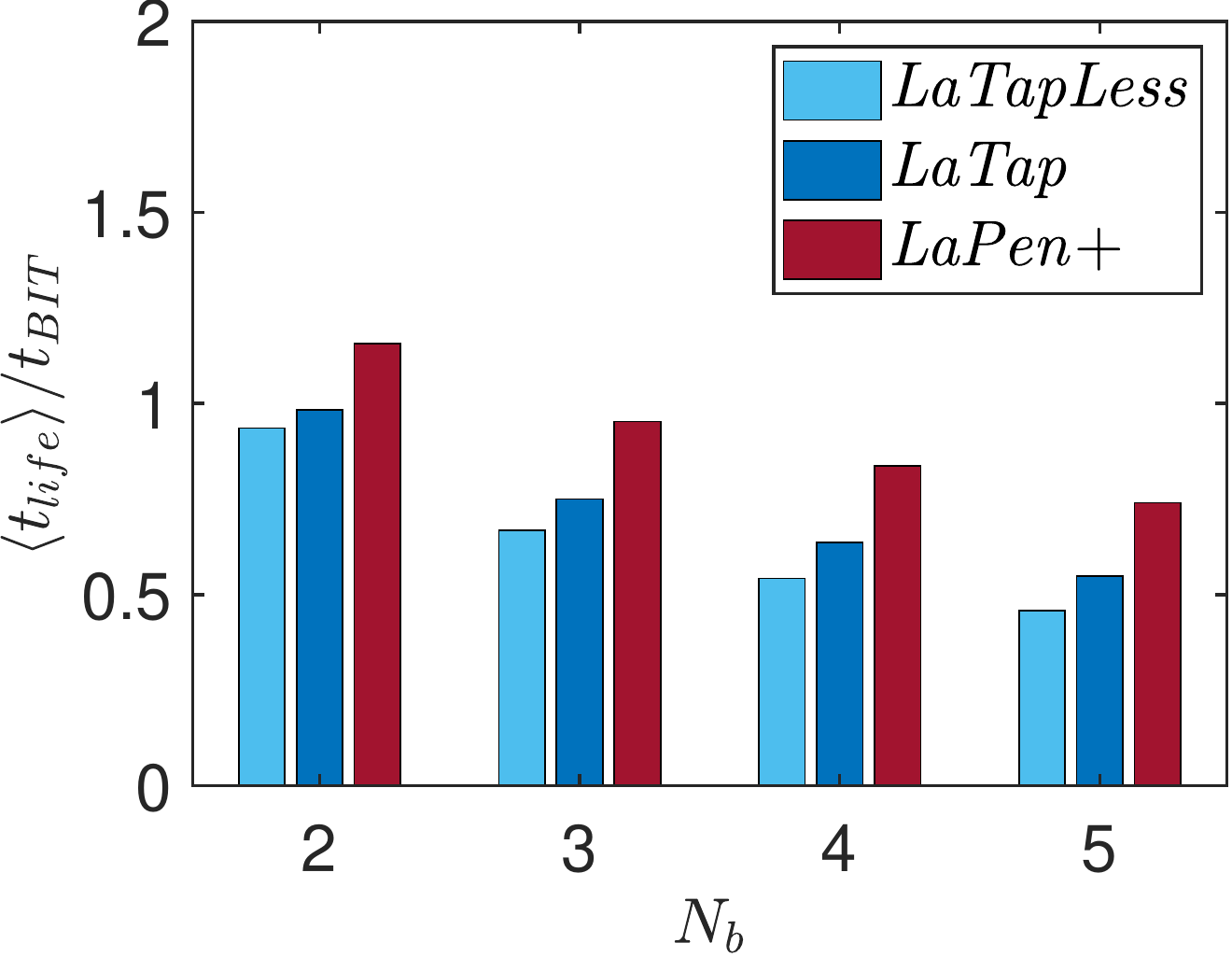}}
	\end{minipage}
	\begin{minipage}[b]{1.0\linewidth}
		\vspace{3mm}
		\centering
		\makebox[0.5em][l]{\raisebox{-\height}{(\textit{c})}}%
		\raisebox{-\height}{\includegraphics[height=3.7cm]{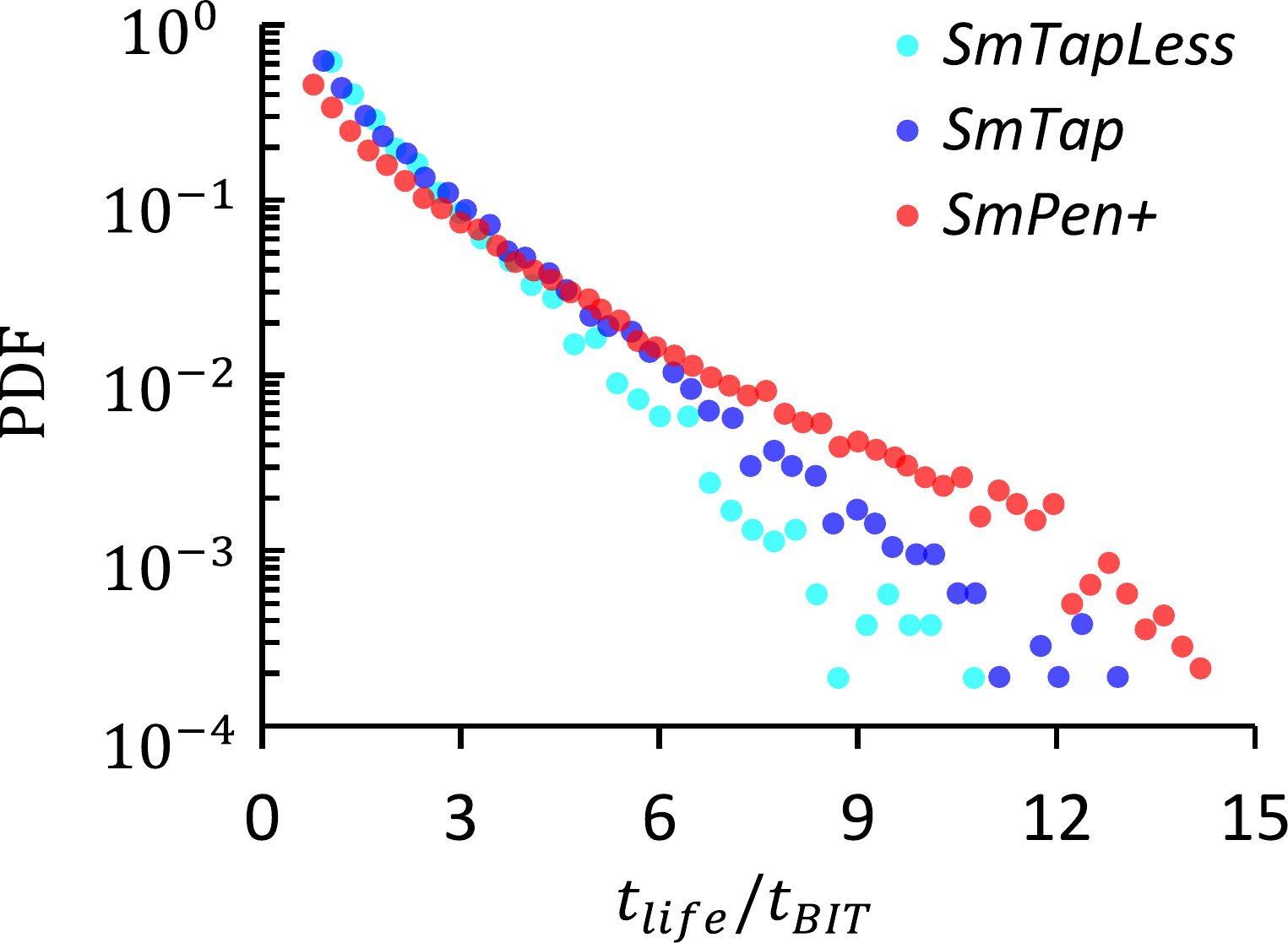}}
        \hspace{5mm}
		\makebox[0.5em][l]{\raisebox{-\height}{(\textit{d})}}%
		\raisebox{-\height}{\includegraphics[height=3.7cm]{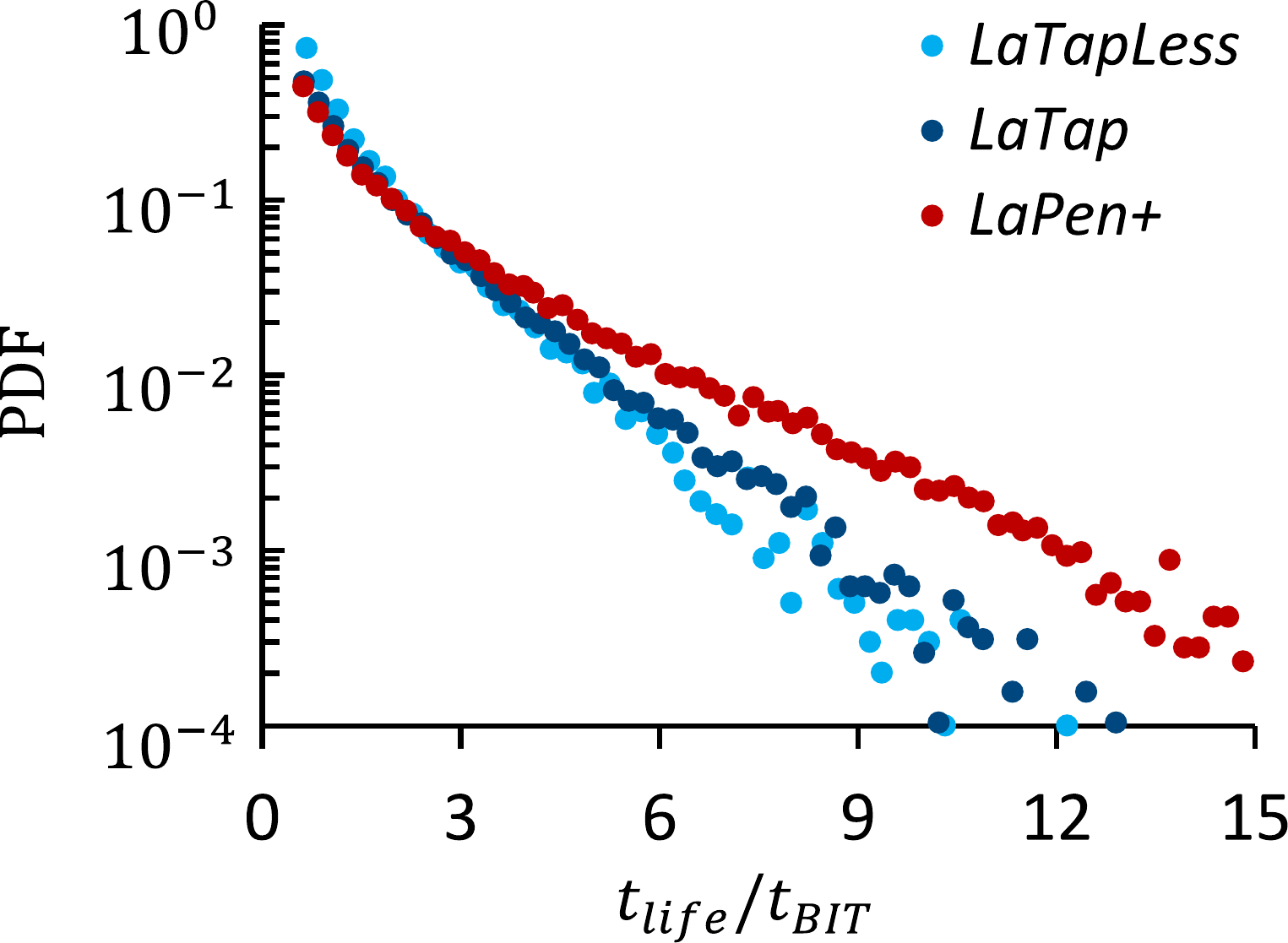}}
	\end{minipage}
\caption{Mean lifetime of 2-, 3-, 4- and 5-bubble clusters (\textit{a,b}) and PDF of the bubble cluster lifetime (\textit{c,d}): the smaller bubble cases (\textit{a,c}) and the larger bubble cases (\textit{b,d}).} \label{fig: lifetime}
\end{figure}

We now turn to consider the mean lifetime of the clusters, $\langle t_{life}\rangle$, as a function of $N_b$ (only the results for $N_b\leq 5$ are shown as the statistics for $N_b>5$ are not converged). Figure \ref{fig: lifetime}(\textit{a,b}) shows  $\langle t_{life}\rangle$ normalized by a characteristic timescale of BIT, $t_{BIT}\equiv d_b/U_b$, where $U_b$ is the mean vertical slip velocity between the bubble and liquid at the column center. The values of $\langle t_{life}\rangle/t_{BIT}$ are order unity, suggesting that $t_{BIT}$ is indeed a dynamically relevant timescale for the cluster lifetimes. The results also reveal a systematic dependence on $N_b$ and the liquid contamination. First, $\langle t_{life}\rangle$ decreases monotonically with increasing $N_b$, such that larger bubble clusters are not only rarer (see \S\,\ref{sec: probability}), but also more unstable. While this may not seem surprising, it is in fact the opposite to what has been observed for inertial particles where the cluster size and its lifetime are positively correlated \citep{2020_Liu}. The difference could be simply due to the fact that the most common values of $N_b$ for our clusters are much smaller than those for the inertial particles in \citet{2020_Liu}, and as a result relatively small changes in the bubble configurations can result in the formation or destruction of a given cluster. The other significant difference is that our bubbles hydrodynamically interact, unlike the numerically simulated inertial particles in \citet{2020_Liu} where a one-way coupling assumption is used. Second, increasing $\alpha$ not only leads to the formation of larger clusters, but also slightly longer mean lifetimes for the clusters, although the lifetimes for $N_b=2$ are the least sensitive to $\alpha$. Third, the mean lifetimes of the bubble clusters notably increase with increasing contamination levels. In a recent paper we showed that increased contamination leads to a reduction of $Re_b$ and an increase in BIT \citep{2023_Ma}. The reduction in $Re_b$ causes the bubble trajectories to be less chaotic, and this may explain why the cluster lifetimes increase with increasing contamination.
  
Figure \ref{fig: lifetime}(\textit{c,d}) show the probability density functions (PDFs) for the cluster lifetimes, which have been computed using clusters of all sizes. The general dependence on the flow variables is similar to that observed for the mean cluster lifetime, with the PDF tails becoming increasingly heavy in the order \textit{TapLess}, \textit{Tap} and \textit{Pen}+ for both the small and large bubbles. The majority of the bubble clusters survive for $t_{life}/t_{BIT}=O(1)$, however, there are extreme cases where clusters survive for up to $t_{life}/t_{BIT}\approx15$ for the \textit{Pen}+ cases. The central regions of the PDFs are well described by stretched exponential functions with parameters that vary between the cases.

\subsection{Mean $N_b$-bubble cluster rise velocity}

\begin{figure}	
	\centering
	\makebox[0.4em][l]{\raisebox{-\height}{(\textit{a})}}
	\raisebox{-\height}{\includegraphics[height=3.75cm]{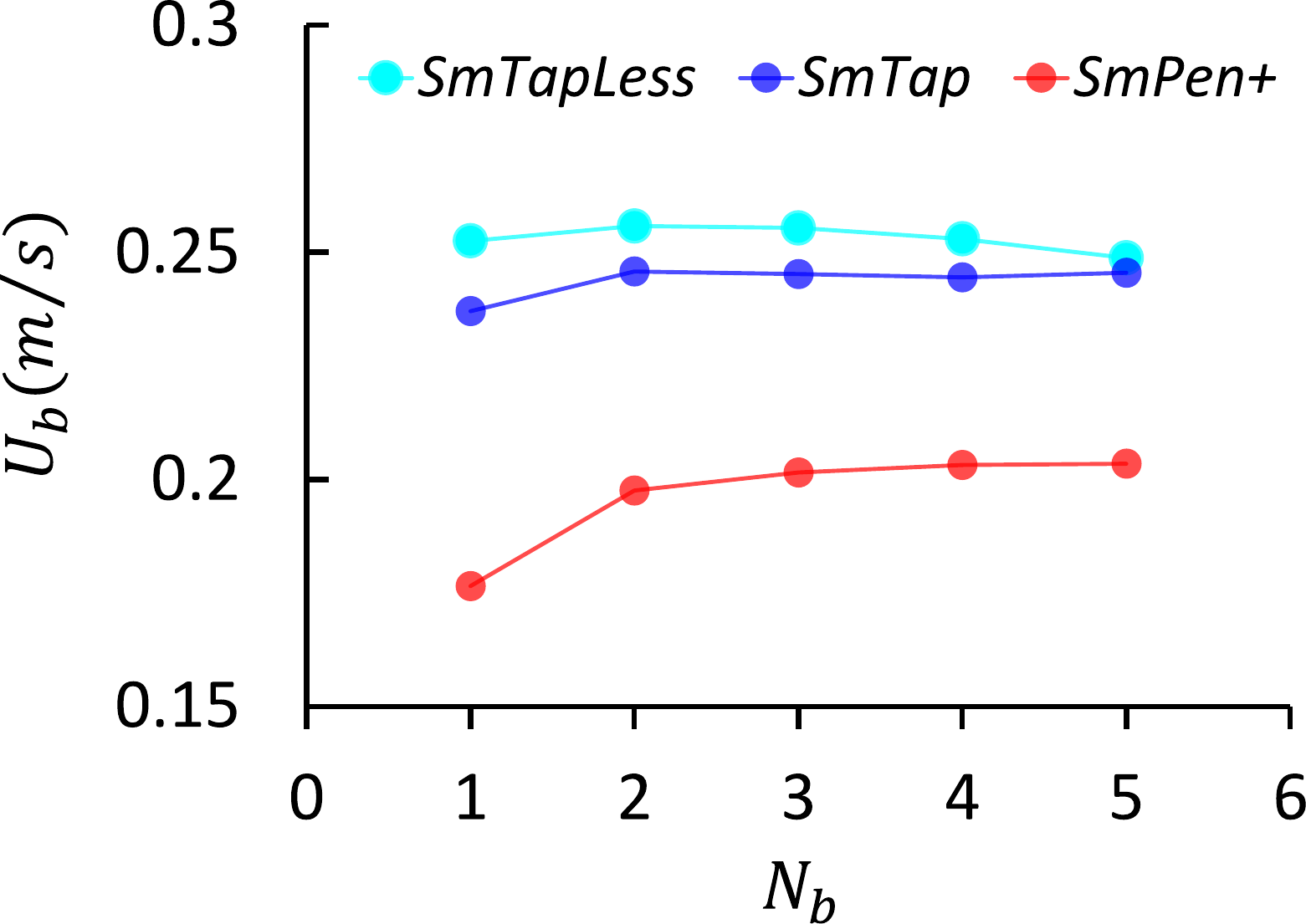}}
	\hspace{5mm}
	\makebox[0.5em][l]{\raisebox{-\height}{(\textit{b})}}
	\raisebox{-\height}{\includegraphics[height=3.75cm]{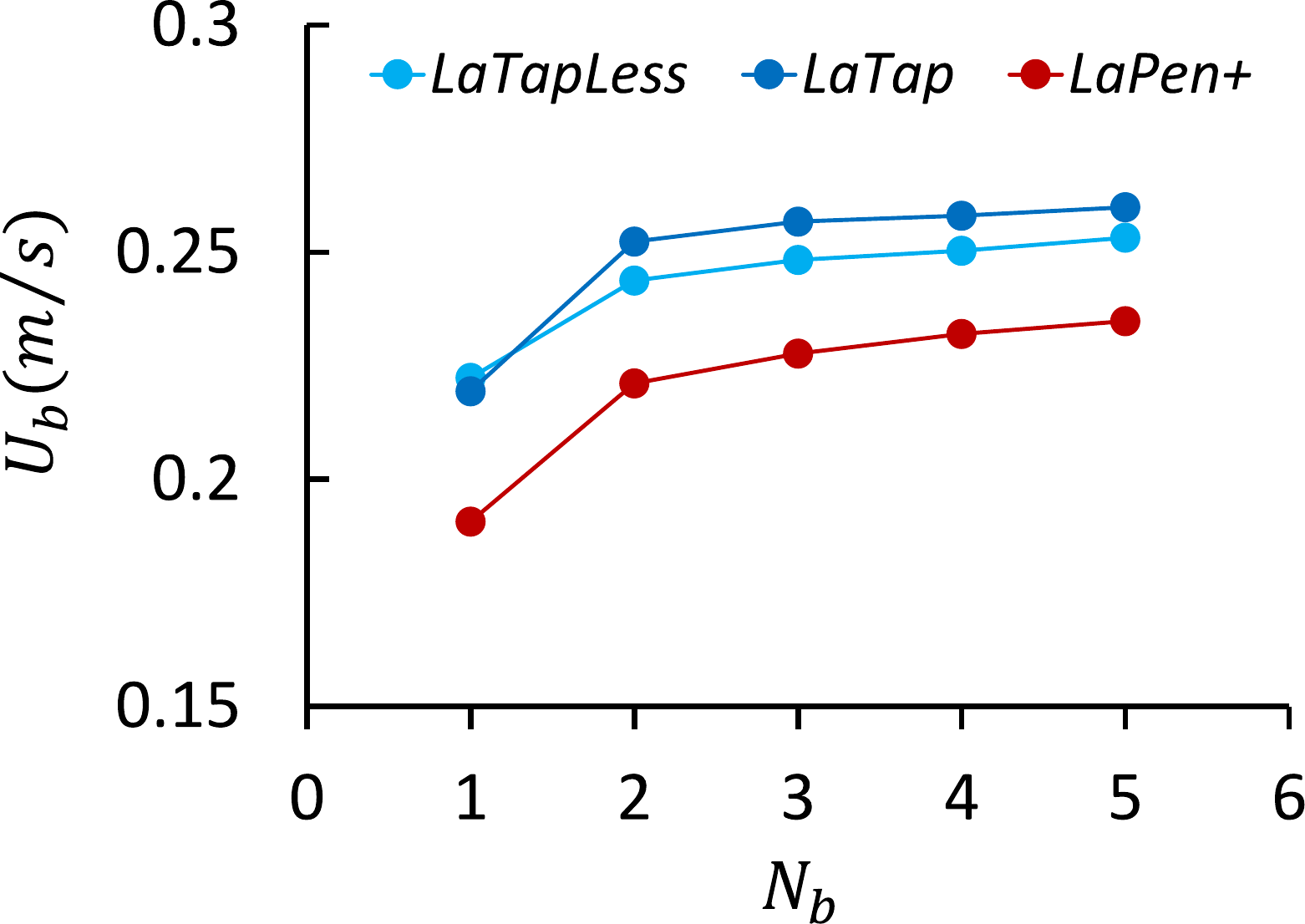}}
	\caption{Mean bubble rise velocity as a function of the number of bubbles $N_b$ in the cluster: (\textit{a}) smaller bubbles and (\textit{b}) larger bubbles. $N_b=1$ denotes unclustered bubbles.} \label{fig: bubble velo}
\end{figure}

We finally consider the role that clustering plays in the mean bubble rise velocity. Figure \ref{fig: bubble velo} shows the mean rise velocity of bubbles in clusters $U_b$ conditioned on $N_b$, and the results for unclustered bubbles $N_b=1$ are also shown for reference. Consistent with our previous results based on averaging over all bubbles \citep{2023_Ma}, the results show that for almost all $N_b$, increasing the liquid contamination leads to a reduction in $U_b$, due to the modification of the bubble boundary conditions. For the larger bubbles we also observe a clear increase in $U_b$ with increasing $N_b$, with an increase of up to $20\%$ when going from unclustered bubbles ($N_b=1$) to bubble pairs ($N_b=2$), while the increase is more moderate when $N_b$ is increased beyond 2. The enhancement of $U_b$ when going from $N_b=1$ to $N_b=2$ is also observed in the \textit{SmPen}+ case, with only slight enhancements when $N_b$ is increased beyond 2. However, for the \textit{SmTapLess} and \textit{SmTap} cases, $U_b$ varies only weakly with $N_b$, even in going from $N_b=1$ to $N_b=2$.

\begin{figure}
	\centering
	\includegraphics[height=6.3cm]{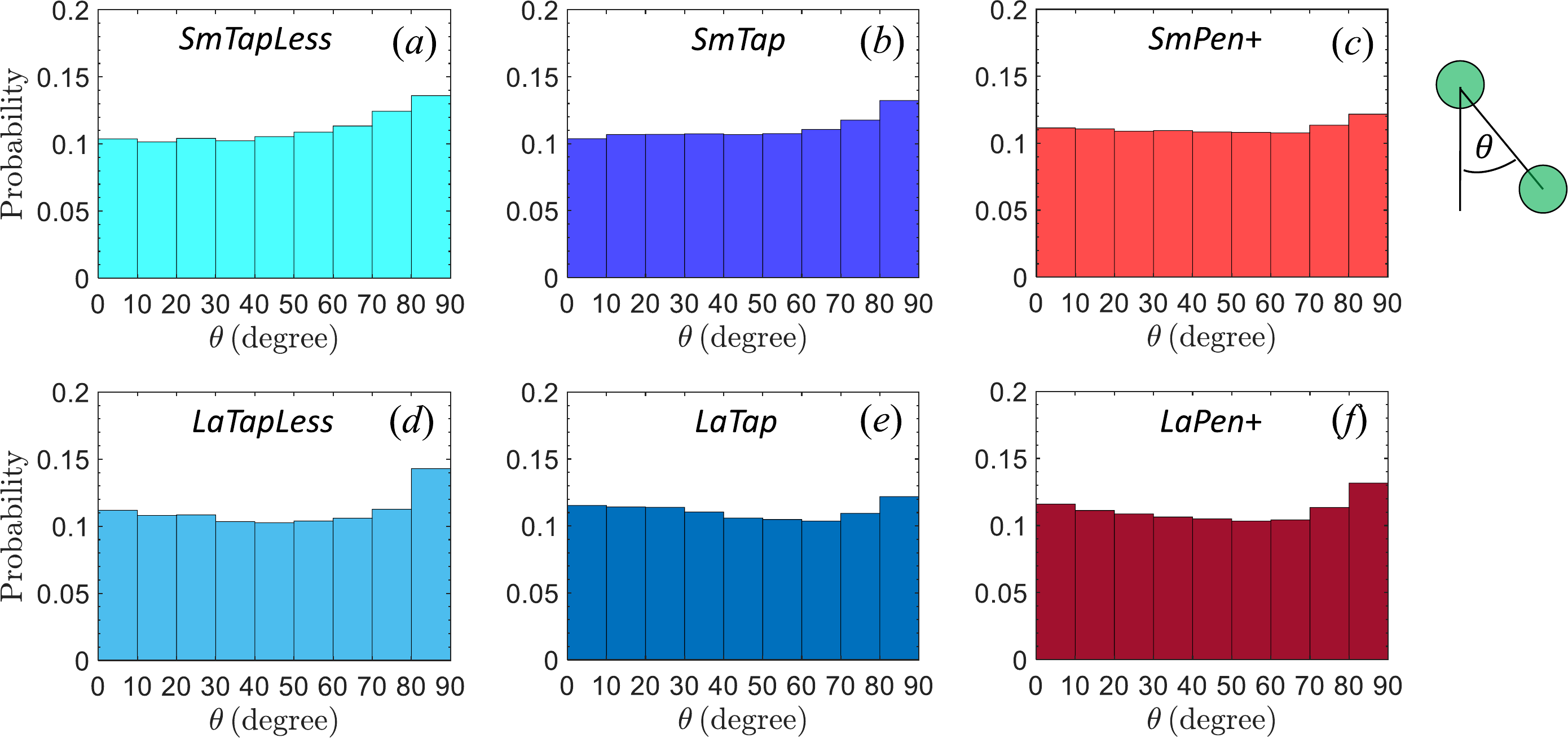}
	\caption{Orientation of bubble pair for the different cases (e.g. in-line bubble pair for $\theta=0^\circ$ and side-by-side for $\theta=90^\circ$).} \label{fig: Pair angle}
\end{figure}

What is the physical explanation for why the clustered bubbles rise faster than unclustered bubbles? We begin by considering the case of bubble pairs $N_b=2$ and plot in figure \ref{fig: Pair angle} the mean inclination angle $\theta$ of bubble pair centreline with respect to the vertical direction (see sketch in the figure). (It should be noted that since our measurements are only quasi-3D, $\theta=0$ does not necessarily mean that the bubbles are in-line because they may nevertheless be separated in the depth direction by up to a distance $2d_b$.) For all cases an almost uniform distribution of $\theta$ is observed, i.e. there is no preferential alignment for bubble pair. This is consistent with visual inspection of the experimental images (see supplementary movies) which show that the bubble pair orientations are not persistent, but instead the bubbles continually trade places in a \textquoteleft leapfrog\textquoteright\ fashion. This observation was also found in many 3D experiments \citep{1995_Stewart,2010_Riboux} and DNS for bubble swarm \citep{2003_Bunner,2005_Esmaeeli}. It is, however, strikingly different from the behaviour observed for isolated bubble pairs where a stable configuration is observed for two clean spherical bubbles where they rise side-by-side \citep{2011_Hallez}, while for contaminated systems their stable configuration is to be in-line due to the lift reversal experienced by the trailing bubble \citep{2022_Atasi}. One possible reason for this discrepancy is that in our experiments the bubbles have oscillating and/or chaotic rising paths, for which the probability that two bubbles will rise in a stable arrangement is very low. By contrast, in \cite{2011_Hallez} the bubbles are fixed at various positions, and in \cite{2022_Atasi} $Re_b$ is small enough such that the bubbles have straight rising paths. Another reason is that in our experiments the bubble pairs are not isolated and can experience fluctuations and turbulence due to the wakes of other bubbles in the flow, and this will readily suppress any preferential orientation that might have occurred were the bubble pairs isolated.

Although the bubble pair orientation is almost random, the impact of their interaction on $U_b$ will however depend on $\theta$, especially in the present bubble regime (deformable bubbles with $Re_b\sim O$(100-1000)).
%Although the bubble pair orientation is almost random, the impact of their interaction on $U_b$ will however depend on $\theta$. For deformable bubbles with $Re_b\sim O(100)$, the effect of interaction is more pronounced on the trailing bubble when it enters an interaction with the wake of the leading bubble.
For example, in the side-by-side configuration the two bubbles are outside of each others wake and the modification to the drag force on each bubble is minimal \citep{2019_Kong}. On the other hand, for the in-line configuration the trailing bubble is sheltered by the leading bubble and the reduced pressure behind the leading bubble causes the trailing bubble to be sucked towards it, increasing the rise velocity of the trailing bubble while the leading bubble is almost unaffected \citep{2021_Zhang}. These effects mean that only the rise velocity of trailing bubbles will be significantly affected by the clustering, and hence when averaged over all orientations, the increased vertical velocity of the trailing bubbles leads to an overall increase in $U_b$. This explains the increased mean rise velocity for $N_b=2$ compared to the $N_b=1$ results in figure \ref{fig: bubble velo}. The increase is, however, minimal for the cases \textit{SmTapLess} and \textit{SmTap}. This is most likely due to the bubble wakes being weaker for these cases, a result of which is that the bubble interaction and the associated effect on $U_b$ is also weaker.

The results in figure \ref{fig: bubble velo} show that $U_b$ further increases as $N_b$ is increased beyond 2. The can be understood in terms of the enhanced opportunity for bubbles to be sheltered by other bubbles as $N_b$ increases. However, the increase is not as strong as when going from $N_b=1$ to $N_b=2$ because the greatest effect of sheltering will occur when two bubbles are in-line; if the in-line bubbles are part of a cluster, the additional bubbles in the cluster must be displaced in the horizontal direction due to the finite size of the bubbles, and they will therefore be less effective in sheltering the trailing bubble. 
It is interesting to note that for experiments on heavy particles settling in a quiescent fluid, similar behaviour was also found, with clustered particles falling faster \citep{2016_Huisman}. In that case, the enhanced settling velocity was also attributed to a sheltering effect, i.e. reduced drag on particles that are falling in the wake of other particles. However, in that context, the particle clusters were found to exhibit strong alignment with the vertical direction, unlike our bubble clusters whose orientations are almost random (at least for the $N_b=2$ case).

\section{Conclusions} 

We conducted experiments on the temporal evolution of bubble clusters with the aid of a new bubble tracking method for crowded swarms. Our results show that 2-bubble clusters are the most common, however, 3 and 4-bubble clusters also often occur. The clusters persist on average for a time of order $d_b/U_b$, although rare clusters persisting for an order of magnitude longer are also observed. Furthermore, surfactants are observed to enhance the cluster sizes and their lifetimes. A positive correlation between cluster size and bubble rise speed is observed, with clustered bubbles rising up to $20\%$ faster than unclustered bubbles. Finally, while our cluster tracking  method is only quasi-3D, a fully 3D method for dense, deformable bubbles can be developed by combining our bubble identification method with the recent tracking algorithm of \cite{2023_Tan} that currently applies to spherical bubbles.

\bibliographystyle{jfm}
\bibliography{Cluster_paper}
\end{document}